\documentclass[aps,prx_,reprint]{revtex4-2}

\usepackage{graphicx}
\usepackage{amsmath,amssymb,amsfonts}
\usepackage{bm} 
\usepackage{hyperref} 
\usepackage{float} 
\usepackage{color}
\usepackage{xcolor}
\usepackage{soul}
\setstcolor{red}
\usepackage{booktabs}
\usepackage{subcaption}
\usepackage[utf8]{inputenc}
\usepackage{multirow}
\usepackage{algorithm}
\usepackage{algorithmicx}
\usepackage{algpseudocode}
\usepackage{listings}
\usepackage{appendix}
\usepackage{caption}
\usepackage{ragged2e}

\makeatletter
 \long\def\@makecaption#1#2{%
   \vskip\abovecaptionskip
   \sbox\@tempboxa{#1. #2}%
   \ifdim \wd\@tempboxa >\hsize
     \begin{minipage}{\hsize}\justifying
       #1. #2\par
     \end{minipage}%
   \else
     \centerline{#1. #2}%
   \fi
   \vskip\belowcaptionskip
 }
\makeatother

\begin{document}

\title[Article Title]{Cyclic Variational Quantum Eigensolver: Escaping Barren Plateaus through Staircase Descent}

\author{Hao Zhang$^{1,2}$, Ayush Asthana$^3$}
\affiliation{$^1$School of Physics and Astronomy, University of Minnesota, Minneapolis, MN 55455, USA}
\affiliation{$^2$Department of Physics, University of Wisconsin – Madison, Madison, Wisconsin 53706, USA}
\affiliation{$^3$Department of Chemistry, University of North Dakota, ND 58202, USA}

\begin{abstract}
We introduce the Cyclic Variational Quantum Eigensolver (CVQE), a hardware-efficient framework for accurate ground-state quantum simulation on noisy intermediate-scale quantum (NISQ) devices. CVQE departs from conventional VQE by incorporating a measurement-driven feedback cycle: Slater determinants with significant sampling probability are iteratively added to the reference superposition, while a fixed entangler (e.g., single-layer UCCSD) is reused throughout. This adaptive reference growth systematically enlarges the variational space in most promising directions, avoiding manual ansatz or operator-pool design, costly searches, and preserving compile-once circuits. The strategy parallels multi-reference methods in quantum chemistry, while remaining fully automated on quantum hardware. Remarkably, CVQE exhibits a distinctive staircase-like descent pattern, where successive energy drops sharply signal efficient escape from barren plateaus. Benchmarks show that CVQE consistently maintains chemical precision across correlation regimes, outperforms fixed UCCSD by several orders of magnitude, and achieves favorable accuracy–cost trade-offs compared to the Selected Configuration Interaction. These results position CVQE as a scalable, interpretable, and resource-efficient paradigm for near-term quantum simulation.
\end{abstract}

\maketitle

\section{Introduction}

The past decade has witnessed rapid progress in quantum hardware, catalyzing the development of a diverse set of practical quantum algorithms tailored for the noisy intermediate-scale quantum (NISQ)\cite{preskillQuantumComputingNISQ2018,bhartiNoisyIntermediatescaleQuantum2022b,arnaultTypologyQuantumAlgorithms2024,kimEvidenceUtilityQuantum2023b, bluvsteinLogicalQuantumProcessor2024, acharyaQuantumErrorCorrection2025, gaoEstablishingNewBenchmark2025, kingBeyondclassicalComputationQuantum2025, gonzalez-cuadraObservationStringBreaking2025}. These approaches already span quantum chemistry chemistry\cite{mcardleQuantumComputationalChemistry2020,asthanaQuantumSelfconsistentEquationofmotion2023,mottaSubspaceMethodsElectronic2024,robledo-morenoChemistryScaleExact2025,xueEfficientAlgorithmsQuantum2025,kaliakinImplicitSolventSampleBased2025}, combinatorial optimization\cite{bauzaScalingAdvantageApproximate2025,zhangComputationalComplexityThreedimensional2025a}, and condensed-matter simulations\cite{babbushLowDepthQuantumSimulation2018,kingBeyondclassicalComputationQuantum2025, yoshiokaKrylovDiagonalizationLarge2025,wangTricriticalKibbleZurekScaling2025,farajollahpourQuantumAlgorithmSoftware2025}.

Hybrid quantum–classical approaches, including the Variational Quantum Eigensolver (VQE)\cite{grimsleyAdaptiveVariationalAlgorithm2019, hugginsNonorthogonalVariationalQuantum2020a,cerezoVariationalQuantumAlgorithms2021a, tangQubitADAPTVQEAdaptiveAlgorithm2021, tillyVariationalQuantumEigensolver2022,fitzpatrickSelfConsistentFieldApproach2024,alvertisClassicalBenchmarksVariational2025}, the Quantum Approximate Optimization Algorithm (QAOA)\cite{farhiQuantumApproximateOptimization2014a, paganoQuantumApproximateOptimization2020,farhiQuantumApproximateOptimization2022,yuQuantumApproximateOptimization2022,zhuAdaptiveQuantumApproximate2022,shaydulinEvidenceScalingAdvantage2024a,munoz-ariasLowdepthCliffordCircuits2024,yuWarmStartAdaptiveBias2025}, and sample-based methods such as Sample-based Quantum Diagonalization (SQD)\cite{robledo-morenoChemistryScaleExact2025,kaliakinImplicitSolventSampleBased2025}, have shown particular promise for near-term applications. In parallel, analog strategies such as Quantum Annealing (QA)\cite{rajakQuantumAnnealingOverview2023,wangManybodyLocalizationEnables2022,zhangCyclicQuantumAnnealing2024,bauzaScalingAdvantageApproximate2025,zhangComputationalComplexityThreedimensional2025a} provide complementary pathways to optimization\cite{bauzaScalingAdvantageApproximate2025,zhangComputationalComplexityThreedimensional2025a}, quantum simulation\cite{kingBeyondclassicalComputationQuantum2025} and quantum neural networks\cite{zhangHowTrainYour2025}, together forming a diverse algorithmic toolbox for the NISQ era.

Among these, the Variational Quantum Eigensolver (VQE) has emerged as one of the leading methods for ground-state quantum simulation. By constructing a parametrized quantum circuit \( U(\boldsymbol{\theta}) \) and minimizing the energy expectation \( \langle \psi(\boldsymbol{\theta}) | H | \psi(\boldsymbol{\theta}) \rangle \) via classical optimization, VQE offers a flexible framework that is naturally robust to noise and decoherence. 

Despite this promise, conventional VQE faces three persistent challenges when applied to hard problems like strongly correlated systems:(i) \emph{Expressivity limits}. Fixed, single-reference ansatz such as the Unitary Coupled Cluster with Singles and Doubles (UCCSD)\cite{cooperBenchmarkStudiesVariational2010,peruzzoVariationalEigenvalueSolver2014a,romeroStrategiesQuantumComputing2018,anandQuantumComputingView2022,fedorovUnitarySelectiveCoupledCluster2022,hirsbrunnerMP2InitializationUnitary2024} fail to capture strong correlation or multi-reference character essential for bond breaking or stretched geometries; (ii) \emph{Optimization Difficulties}. Barren plateaus\cite{mcardleQuantumComputationalChemistry2020,laroccaBarrenPlateausVariational2025a,wierichsAvoidingLocalMinima2020} and rugged landscapes stall parameter updates, particularly as the number of variational parameters increases; and (iii) \emph{Resource overhead}. Achieving a small energy error to the exact ground state or chemical accuracy often requires large circuits, extensive measurements, and long coherence times, straining current NISQ hardware.

\begin{figure*}[ht]
    \centering
    \includegraphics[width=1\linewidth]{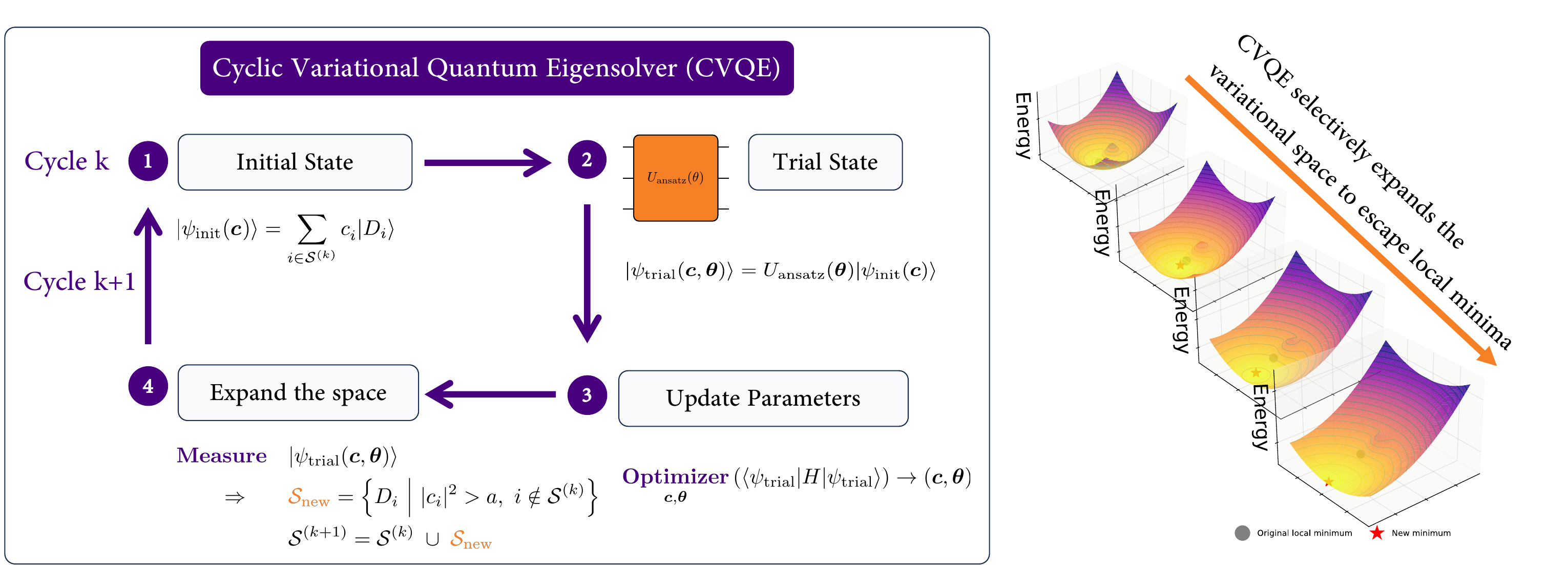}
    \caption{
    \textbf{Schematic workflow of the Cyclic Variational Quantum Eigensolver (CVQE).} Each cycle begins with a reference state composed of selected Slater determinants. The fixed entangling ansatz $U_{\mathrm{UCCSD}}(\boldsymbol{\theta})$ is applied to produce a variational trial state, whose energy is minimized with respect to both the ansatz parameters $\boldsymbol{\theta}$ and the determinant coefficients $\mathbf{c}$. The optimized state is then sampled in the computational basis to identify additional determinants with probabilities above a threshold $a$, which are incorporated into the reference for the next cycle. This feedback loop systematically expands the accessible Hilbert space in the most promising directions. The right panel illustrates how CVQE systematically escapes local minima by opening new directions in the variational space.
    }
    \label{fig:cvqe}
\end{figure*}

Extensions such as ADAPT-VQE\cite{grimsleyAdaptiveVariationalAlgorithm2019,tangQubitADAPTVQEAdaptiveAlgorithm2021,anastasiouTETRISADAPTVQEAdaptiveAlgorithm2024} partially address these limitations by adaptively growing the ansatz. However, they do so at the cost of system-specific circuits, manual operator-pool design and expensive operator evaluations. This leaves open the central question: how can we systematically enlarge the variational space while preserving efficiency and stability on NISQ devices?

In this work, we introduce the \emph{Cyclic Variational Quantum Eigensolver} (CVQE), a hardware-efficient, auto-adaptive framework that establishes a new paradigm for variational quantum simulation. Rather than expanding the ansatz itself, CVQE grows the reference state adaptively through measurement-driven feedback: after each optimization cycle, Slater determinants with high sampling probability are incorporated into the next cycle’s reference superposition, while the entangling structure remains fixed as a fixed ansatz like single-layer UCCSD circuit. This strategy systematically expands the accessible Hilbert space in the most promising directions without manual ansatz or operator pool design, while preserving compile-once, hardware-friendly circuits.

A distinctive hallmark of CVQE is its \emph{staircase descent pattern}: extended energy plateaus are punctuated by sharp downward steps when new determinants are incorporated and the optimizer re-explores freshly opened directions. This behavior arises naturally from CVQE’s cyclic expansion of the reference space, which continuously reshapes the optimization landscape and creates new opportunities for progress. Unlike conventional VQE, where convergence often stalls in barren plateaus, CVQE repeatedly unlocks steep descent paths that drive the energy toward the ground state. To support fresh re-exploration while accelerating convergence, we introduce a Cyclic Adamax (CAD) optimizer. CAD leverages momentum to speed up parameter updates, but periodically resets its momentum variables to adapt to the newly expanded energy landscape. This design amplifies the staircase descent pattern of CVQE, enabling efficient escapes from plateaus and delivering accuracies several orders of magnitude beyond those of fixed UCCSD.

We benchmark CVQE on molecular dissociation problems spanning weakly to strongly correlated regimes: BeH$_2$, H$_6$, and N$_2$. Across all bond lengths, CVQE consistently attains chemical accuracy, converges reliably via its cyclic feedback mechanism, and does so with only a single UCCSD layer. Comparisons with selected configuration interaction methods, including state-of-the-art semistochastic heat-bath Configuration Interaction (SHCI)\cite{holmesHeatBathConfigurationInteraction2016,sharmaSemistochasticHeatBathConfiguration2017,liFastSemistochasticHeatbath2018}, show that CVQE achieves these accuracies with fewer determinants than required, highlighting favorable accuracy–cost trade-offs that are especially advantageous for NISQ devices. 

CVQE exemplifies a \emph{“compile once, optimize everywhere”} philosophy: A fixed ansatz can be deeply optimized on the hardware side and reused across many different applications. The cyclic measurement-driven feedback mechanism enables CVQE to expand the optimization space automatically to the most promising directions, bypassing barren plateaus, maintaining chemical accuracy across weakly and strongly correlated regimes, and delivering determinant efficiency that surpasses advanced classical approaches. This opens a scalable pathway for accurate quantum simulation on near-term hardware.

\section{Main idea}

Cyclic Variational Quantum Eigensolver (CVQE) is a dynamically evolving quantum variational algorithm. It has a measurement-driven feedback cycle mechanism that adaptively expands the variational space to escape local minima and barren plateau; see the schematic diagram in Fig.~\ref{fig:cvqe}.  

In conventional VQE, a parametrized trial state
\begin{equation}
|\psi(\boldsymbol{\theta})\rangle = U(\boldsymbol{\theta})|\psi_{\mathrm{init}}\rangle
\end{equation}
is optimized to minimize the energy 
\begin{equation}
E = \langle \psi(\boldsymbol{\theta})|H|\psi(\boldsymbol{\theta})\rangle.     
\end{equation}
 
A common choice for \(U(\theta)\) in quantum chemistry is the UCCSD ansatz:
\begin{equation}
U(\theta) = e^{T(\boldsymbol{\theta}) - T^\dagger(\boldsymbol{\theta})}.
\end{equation}
Here, the excitation operator \(T = T_1 + T_2\) consists of single (\(T_1\)) and double (\(T_2\)) excitations from occupied to virtual orbitals, parameterized by angles \(\boldsymbol{\theta}\). While effective in many scenarios, UCCSD is inherently limited when describing strongly correlated systems or multi-reference character, owing to its fixed, single-reference nature.

CVQE addresses this limitation by iterating through four key steps in each cycle:
\vspace{0.3em}

\noindent\textbf{(1) Initial state preparation (cycle \(k\)).}  
The initial reference state is a linear combination of selected Slater determinants $|D_i\rangle$ from previous cycles (set $\mathcal{S}^{(k)}$):

\begin{equation}
|\psi_{\mathrm{init}}^{(k)}(\mathbf{c})\rangle = \sum_{i\in\mathcal{S}^{(k)}} c_i |D_i\rangle.
\end{equation}

The coefficients of newly identified determinants are initialized with small values whose magnitude is scaled relative to the global gradient norm of coefficients, thereby adapting naturally to the optimization stage (see Methods), promoting exploration without destabilizing convergence.
In the first cycle, $\mathcal{S}^{(1)}$ is usually chosen as the Hartree-Fock state $\{|\mathrm{HF}\rangle\}$ for molecules.
\vspace{0.3em}

\noindent\textbf{(2) Trial state preparation.}  
A fixed entangling unitary \(U_{\mathrm{ansatz}}(\boldsymbol{\theta})\) (UCCSD in this work) acts on the initial reference state to produce the trial state:

\begin{equation}
|\psi_{\mathrm{trial}}(\mathbf{c}, \boldsymbol{\theta})\rangle = U_{\mathrm{ansatz}}(\boldsymbol{\theta})\,|\psi_{\mathrm{init}}^{(k)}(\mathbf{c})\rangle.
\end{equation}

Thus, the optimization simultaneously targets both the coefficients \(\mathbf{c}\) and the unitary parameters \(\boldsymbol{\theta}\). 
\vspace{0.3em}

\noindent\textbf{(3) Parameter update.}  
Both \(\mathbf{c}\) and \(\boldsymbol{\theta}\) are optimized using the classical optimizers, once in each cycle.
\begin{equation}
    \underset{\boldsymbol{c}, \boldsymbol{\theta}}{{\text{Optimizer}}}
\left( \langle \psi_{\text{trial}} | H | \psi_{\text{trial}} \rangle \right)
\to (\boldsymbol{c}, \boldsymbol{\theta})
\end{equation}
In this work, we use gradient descent to optimize $\boldsymbol{\theta}$ and CAD optimizer to optimize $\mathbf{c}$. 
\vspace{0.3em}

\noindent\textbf{(4) Space expansion.}  
The optimized trial state is sampled in the computational basis. New Slater determinants with measured probability \(|c_i|^2\) above a threshold are added to the determinants set \(\mathcal{S}^{(k+1)}\). 
This measurement-driven reference growth allows CVQE to bypass key limitations of conventional variational approaches: by adaptively enriching the reference state, the algorithm continually opens new optimization pathways, solving issues such as barren plateaus where gradient magnitudes vanish. Unlike other adaptive schemes, such as ADAPT-VQE, which require costly expectation evaluations over large operator pools, CVQE identifies promising directions directly from measurement statistics. As a result, it costs only a small number of measurement shots to discover new variational degrees of freedom. This cyclic variational process adaptively reshapes the optimization landscape and refines the trial state, enabling CVQE to systematically escape local minima and barren plateaus (as illustrated in the right panel of Fig.~\ref{fig:cvqe}), progressively improving its fidelity to the true ground state. 
With the measurement-driven feedback cycle, it opens the variational space only in important directions, maintaining a relatively shallow circuit, fixed ansatz throughout the calculation.

The adaptive scheme proposed also resembles the developments in multi-reference methods in quantum chemistry, where a distinction is made in the molecular electronic correlation between dynamical and non-dynamical correlations. The non-dynamical part of the correlation is usually targeted by changing the reference state of Hartree-Fock to a collection of useful reference configurations. The choice of these reference configurations is not trivial and is based on chemical intuition. Complete active space and restricted active space are popular choices of developing a list of these configurations in quantum chemistry \cite{keller2015selection}. CVQE, on the other hand, uses an automated strategy to build a list of these configurations. In classical quantum chemistry methods, once the list of reference configurations are built, it is followed by a method to treat the non-dynamical correlation. Several methods have been historically used for this step, for example, Perturbation theory based MRPT2 and NEVPT2~\cite{schreiber2008benchmarks,guo2021approximations} or coupled cluster based IC-MRCC and Mk-MRCC~\cite{das2008development,evangelista2018perspective} methods. The automated reference building strategy in CVQE can also be followed by various quantum computing methods to treat non-dynamical correlation, including UCCSD, ADAPT-VQE and other similar quantum algorithms.

\begin{figure*}[htbp]
    \centering
    \includegraphics[width=1\linewidth]{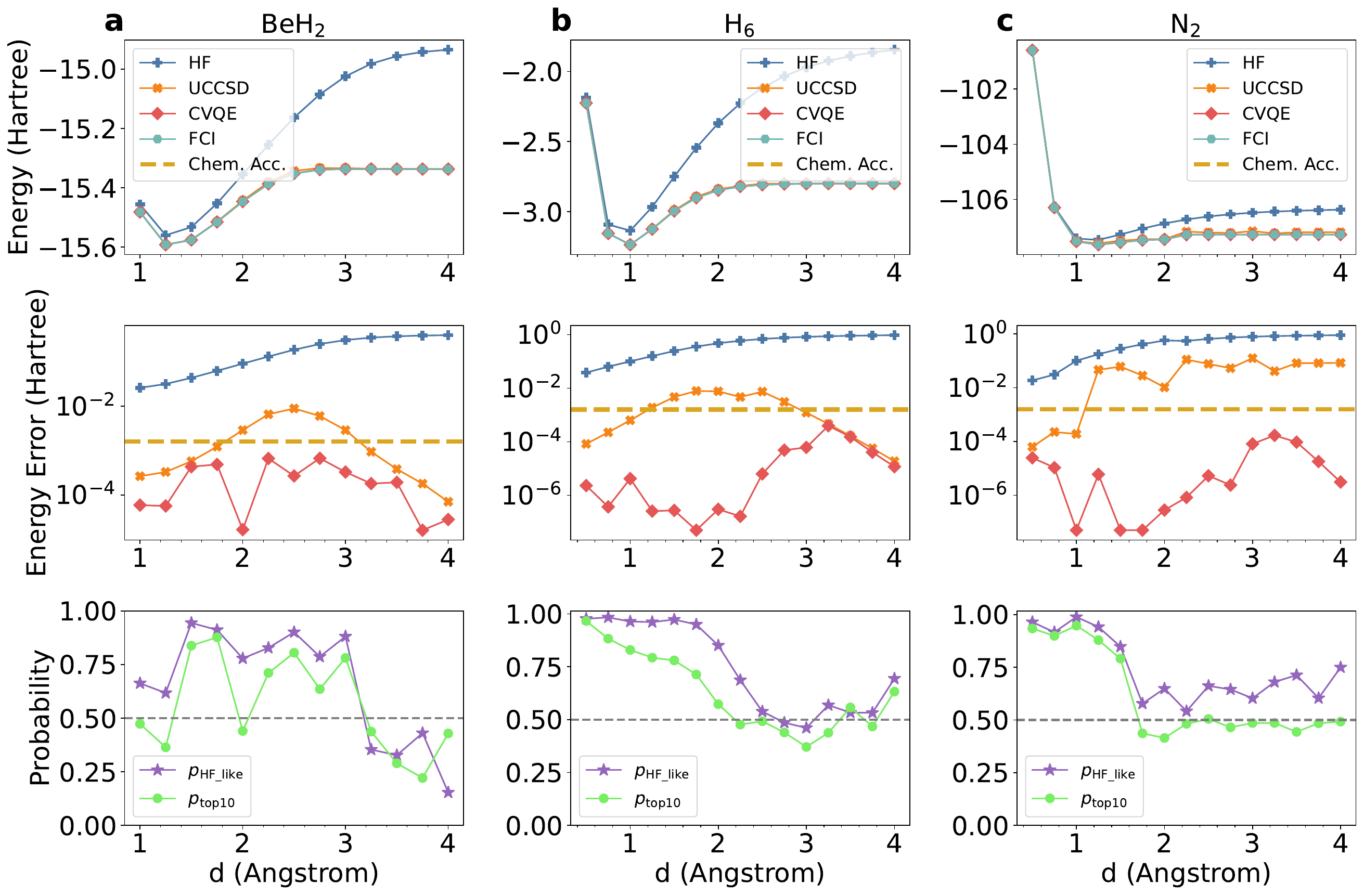}
    \caption{
    \textbf{Performance of CVQE across dissociation profiles of (a) BeH$_2$, (b) H$_6$, and (c) N$_2$.} The top panels show potential energy surface relative to full configuration interaction (FCI); the middle panels present absolute energy errors relative to FCI on a logarithmic scale; the bottom panels report diagnostic measures from the final optimized CVQE wavefunction: $p_{\mathrm{HF\_like}}$ (cumulative probability weight of the HF-like determinants) and $p_{\mathrm{top10}}$ (cumulative probability weight of the ten most probable determinants). CVQE consistently achieves chemical accuracy ($1.6\times 10^{-3}$ Ha) across weakly and strongly correlated regimes, even when UCCSD errors increase by orders of magnitude.
    }
    \label{fig:cvqe_accuracy}
\end{figure*}

\section{Results}

\subsection{Molecular dissociation and accuracy across correlation regimes}
\label{subsec:results_dissociation}
A key task for any variational quantum algorithm is its ability to maintain accuracy across different correlation regimes, particularly in the challenging multi-reference problems that arise in bond-breaking. To evaluate the performance and generality of CVQE, we benchmarked it on three representative molecular systems, BeH\textsubscript{2}, H\textsubscript{6}, and N\textsubscript{2}, across a range of bond distances. These systems were selected because they span diverse correlation regimes and feature well known challenging electronic structures, while still remaining tractable for quantum algorithm simulations.

BeH\textsubscript{2} serves as a prototypical small system where stretching the Be–H bonds induces moderate static correlation, exposing the limitations of single-reference methods without overwhelming the Hilbert space size. H\textsubscript{6} in a linear chain configuration is a canonical strongly correlated testbed: as H–H distances increase, the system transitions from a delocalized to a localized regime, leading to near-degeneracies between many Slater determinants. N\textsubscript{2} challenges variational ansatz with its triple bond, where bond dissociation demands a simultaneous treatment of strong static and dynamic correlations.

All simulations were carried out within the Born–Oppenheimer approximation using the STO-3G basis set. The electronic Hamiltonians were formulated in second quantization in the molecular orbital basis and subsequently mapped to qubit Hamiltonians via the Jordan–Wigner transformation.

Figure~\ref{fig:cvqe_accuracy} presents (i) the potential energy surfaces (PES) compared to full configuration interaction (FCI), see top row, (ii) the energy errors relative to FCI on a logarithmic scale, see middle row, (iii) diagnostic information about optimized reference state. CVQE energies is compared with Hartree-Fock energies, UCCSD energies, chemical accuracy  ($1.6\times10^{-3}$ Ha) and FCI energies.
Across all systems and bond lengths, CVQE maintains remarkably stable within the chemical accuracy threshold, even where UCCSD errors grow by orders of magnitude. For BeH$_2$ and H$_6$, UCCSD errors grow to $10^{-2}$ Ha in stretched regimes, while CVQE remains at $\mathcal{O}(10^{-6}$--$10^{-3})$ Ha. For N$_2$, UCCSD deviates significantly beyond $d\gtrsim 1.25~\text{\AA}$, but CVQE follows FCI within chemical accuracy across the curve.

The diagnostic probabilities shed light on the underlying mechanism. The HF-like weight $p_{\mathrm{HF\_like}}$, see purple lines, is defined as the cumulative coefficient norm $|c_i|^2$ of determinants closely related to Hartree-Fock state (including up to double excitations) within the final optimized reference. This weight is large near equilibrium but decreases as bond are stretched, reflecting the diminishing dominance of the Hartree–Fock determinant. To keep within the chemical accuracy, CVQE manages to add more diverse set of determinants in the reference state. On the other hand, the cumulative coefficient norm of the ten most probable determinants $p_{\mathrm{top10}}$ remains between $0.2$ and $0.8$ across distances, indicating that contributions are unevenly distributed and dominated by a compact subset. This compressibility demonstrates that CVQE can retain accuracy with a relatively small determinant set, without substantially increasing circuit depth.

\subsection{Convergence dynamics}
\label{subsec:results_training}

\begin{figure*}[htbp]
    \centering
    \includegraphics[width=1\linewidth]{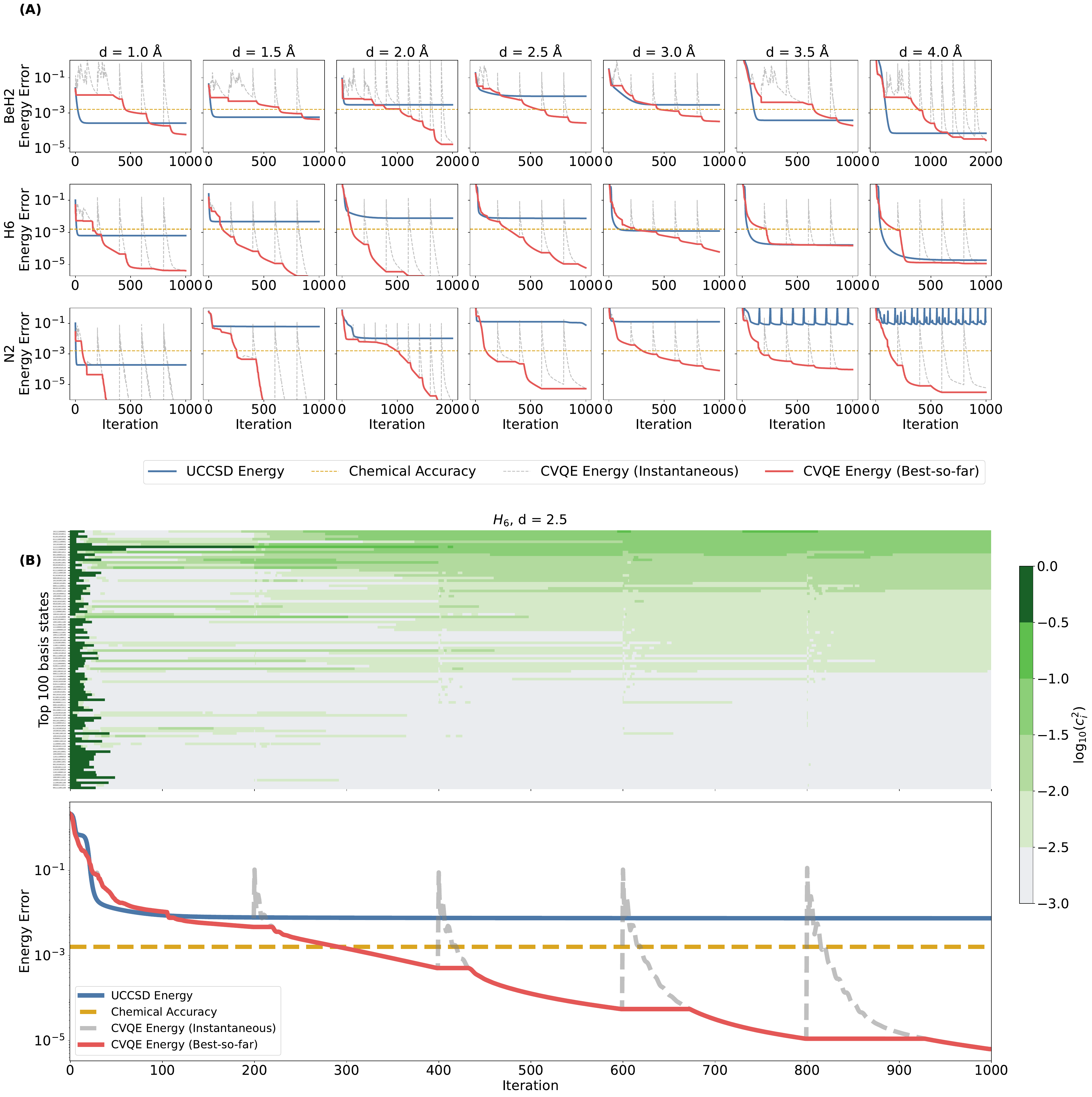}
    \caption{
    \textbf{Convergence dynamics of CVQE versus UCCSD.}
    (A) Best-so-far energy error across bond lengths for BeH$_2$, H$_6$, and N$_2$. UCCSD (blue) quickly plateaus, while CVQE (red/grey) exhibits a staircase-like convergence: sharp energy drops arise when new determinants are added and the optimizer (CAD) resets momentum, enabling escape from barren plateaus.
    (B) Detailed view for H$6$ at $d=2.5$ Å. Top panel tracks logarithmic weights $\log_{10}|c_i|^2$ of the 100 most significant determinants, showing sequential emergence and amplification of key configurations while less relevant ones decay. Bottom panel: CVQE continues to reduce energy well past the UCCSD plateau, achieving chemical accuracy by a wide margin.
    }
    \label{fig:energy_curve}
\end{figure*}

While static benchmarks establish the accuracy of CVQE, understanding its convergence behavior is essential for evaluating its scalability and optimization stability. To this end, we compared the optimization dynamics of CVQE with UCCSD across a range of bond lengths for each molecule, see Fig.~\ref{fig:energy_curve}A.

UCCSD (blue curves) typically exhibits a rapid initial decrease in error, but then stalls at relatively high values, reflecting the limited expressivity of a fixed single-reference ansatz. In contrast, CVQE (red curves) continues to improve far beyond this plateau. Its trajectory displays a distinctive staircase profile: flat regions of slow progress punctuated by abrupt drops in energy error. These sharp decreases occur whenever two events coincide: (i) new determinants are incorporated into the reference superposition, expanding the variational space, and (ii) the optimizer resets its momentum, enabling exploration of newly opened directions. The grey dashed line in Fig.~\ref{fig:energy_curve}A tracks the instantaneous CVQE energies during iterations. After momentum is reset every 200 steps, the optimizer rapidly escapes from the previous plateau and discovers more favorable descent directions. This robust pattern across all tested systems underscores the stability and scalability of CVQE: by cyclically reshaping the optimization landscape and redirecting trajectories toward favorable descent paths, the algorithm consistently avoids stagnation in local minima or barren plateaus.

A more detailed view is shown for H$_6$ at $d=2.5~\text{\AA}$ in Fig.~\ref{fig:energy_curve}B. The upper panel tracks the logarithm weights $\log_{10}|c_i|^2$ of the top 100 determinants throughout optimization. The color scale encodes the determinant weights, where darker colors correspond to larger values. Determinants compete with one another upon entering the reference, with important ones being progressively amplified while less relevant ones gradually decay. At different stages, new significant determinants emerge sequentially, opening fresh directions for the optimization to explore.

In the lower panel, after UCCSD stalls above the chemical-accuracy threshold, CVQE continues to reduce the energy and ultimately surpasses chemical accuracy by a wide margin. This behavior directly reflects the determinant dynamics shown in the upper panel: as low-weight states fade and new important determinants emerge, the optimization is adaptively redirected. Together, these features demonstrate how cyclic reference expansion systematically reshapes the energy landscape and prevents the optimizer from being trapped in local minima or barren plateaus.

\subsection{Accuracy-cost control under determinant budget constraints}
\label{subsec:results_ndets}

A central advantage of CVQE is its explicit control over the number of determinants, which directly regulates both measurement cost and circuit complexity (see Methods). This tunability is particularly valuable for near-term hardware, where shot budgets and coherence times are limited. Once the determinant budget is reached, newly identified promising configurations replace those with the smallest coefficients, ensuring that the retained set remains compact yet effective.

\begin{figure}[htbp]
    \centering
    \includegraphics[width=1\linewidth]{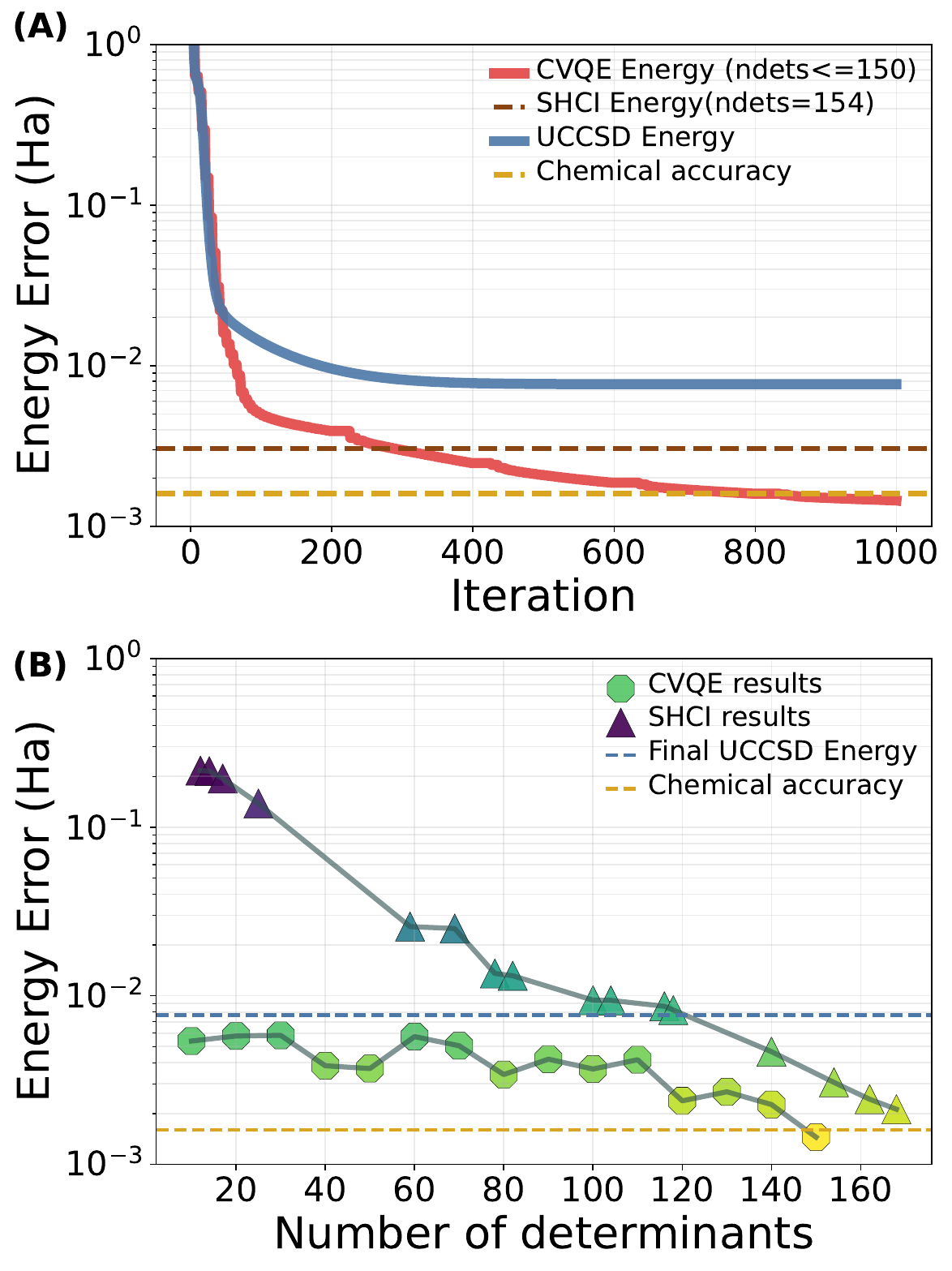}
    \caption{
    \textbf{Accuracy–cost trade-off under determinant budget constraints for H$_6$ at $d=2.0~\text{\AA}$.} 
    (A) Representative run with $n_{\mathrm{dets}} \le 150$. CVQE (red) reduces the errors to the $10^{-3}$ Ha regime, one order of magnitude below UCCSD (blue), and even surpasses SHCI (black) of comparable size ($n_{\mathrm{dets}} = 154$) in early iterations.
    (B) Aggregate results with varying determinant budgets. Both CVQE and SHCI improve as $n_{\mathrm{dets}}$ increases, but CVQE consistently achieves lower errors across all tested budgets.
    }
    \label{fig:ndets}
\end{figure}

To quantify this accuracy-cost control, we simulated the H$_6$ chain at strong correlation, $d=2.0~\text{\AA}$, under different determinant caps $n_{\mathrm{dets}}$. Figure~\ref{fig:ndets} summarizes these results. In panel A ($n_{\mathrm{dets}}\le150$), CVQE reduces the energy error into the $10^{-3}$ Ha regime, while UCCSD remains an order of magnitude less accurate. Notably, CVQE already surpasses a state-of-the-art selected CI method (SHCI)\cite{holmesHeatBathConfigurationInteraction2016,sharmaSemistochasticHeatBathConfiguration2017,liFastSemistochasticHeatbath2018} with a comparable determinant count $n_{\mathrm{dets}}=154$ during the early iterations ($\sim$200–300). Figure~\ref{fig:ndets}B aggregates results over varying $n_{\mathrm{dets}}$, showing that errors decrease steadily with budget size for both CVQE and SHCI. However, across all tested budgets, CVQE consistently outperforms SHCI.

These benchmarks highlight the high determinant efficiency of of CVQE: chemical accuracy is typically achieved with determinants fewer than required by SHCI. This built-in tunability of the determinant budget also offers a practical knob for balancing accuracy against hardware resources, making CVQE exceptionally well-suited for NISQ-era devices.

\section{Discussion}

The Cyclic Variational Quantum Eigensolver (CVQE) establishes a new paradigm for NISQ-era quantum simulation by combining adaptive reference growth with a fixed entangler. This framework consistently achieves chemical accuracy across both weakly and strongly correlated regimes, while avoiding the need for manual design and costly search of ansatz.

\vspace{0.3em}
\textbf{High Expressivity through Feedback.}

By cyclically sampling the quantum state and incorporating important Slater determinants into the reference, CVQE systematically expands its variational space along promising directions. This measurement-driven feedback strategy circumvents the expressivity bottleneck of fixed single-reference ansatze such as UCCSD, while avoiding the expensive operator-pool scans required in adaptive approaches like ADAPT-VQE. The result is a lightweight and automated mechanism for opening optimization pathways that deliver consistent chemical accuracy.

\vspace{0.3em}
\textbf{Linear Cost with Determinant Control.}

Relative to plain UCCSD, CVQE introduces moderate but scalable overheads. In our experiments, preparing the reference superposition requires one auxiliary qubit and some additional gates, which scale linearly in the number of retained determinants. More generally, with $m$ ancillary qubits, preparing a superposition of $d$ determinants over $n$ qubits has been shown to scale linearly as
$O\left(\tfrac{n}{\log (n+m)}d+n\right)$\cite{liNearlyOptimalCircuit2025}. If the specific structure of the low-energy subspace is exploited, the cost could be reduced even further, leading to more efficient preparation schemes.
 
Optimizing determinant coefficients increases the number of variational parameters and hence the required measurement shots, which also grow linearly with determinant count. Crucially, the determinant budget $n_\text{dets}$ provides a tunable knob to balance accuracy against cost. Together, these features make render the overhead of CVQE linear, affordable and adjustable, well aligned with the constraints of NISQ hardware.

\vspace{0.3em}
\textbf{Staircase Descent to Escape Plateaus.}

The hallmark staircase trajectory, flat plateaus punctuated by sharp energy drops, arises from the interplay between cyclic reference expansion and the CAD optimizer restarts. Each restart realigns the optimizer to the enlarged landscape, systematically unlocking new descent directions and enabling escape from barren plateaus or local minima. This robust mechanism provides a practical, hardware-friendly route to overcoming one of the central challenges in variational quantum algorithms.

\vspace{0.3em}
\textbf{Compile Once, Optimize Everywhere.}

Unlike adaptive ansatz methods whose circuits grow system-specific, CVQE employs a fixed entangler throughout. This design enables the entangling circuit to be compiled, calibrated, and deeply optimized at the hardware level only once. Once optimized, the same entangler can be reused across a wide range of molecular systems, correlation regimes, and even different applications, without any need for re-engineering. Such a design dramatically reduces compilation and calibration cost, since the expensive hardware-level procedures no longer need to be repeated for each instance.

Equally important, a single global calibration can be performed to suppress systematic errors in the entangler. Because the circuit structure remains fixed, hardware-level optimization of gate implementations can be carried out with much greater effectiveness, reducing error rates and crosstalk more significantly than would be possible for continuously changing adaptive circuits. As a result, CVQE becomes more robust and accurate on realistic quantum devices.

Finally, the universality of the compiled entangler suggests the possibility of developing specialized high-performance quantum devices for quantum chemistry and materials simulation. By tailoring a hardware-efficient entangler to chemistry and material workloads, one could design dedicated platforms where the same deeply optimized circuit supports a broad class of problems. This makes CVQE not only resource-efficient but also a practically attractive approach for deployment on near-term quantum hardware.

\vspace{0.3em}
\textbf{Physical and Chemical interpretability.}

A distinctive advantage of CVQE lies in the interpretability of its selected determinants, which sets it apart from purely parameterized ansatze. The determinants incorporated into the reference state correspond to physically meaningful electronic configurations, thereby offering not only quantitative accuracy but also qualitative chemical insight. For instance, during N$_2$ dissociation, the emergence of additional determinants directly reflects bond-breaking configurations, establishing a transparent connection between the algorithm’s internal dynamics and chemically relevant processes.

\vspace{0.3em}
\textbf{Future directions.}

Several avenues remain for further development. First, determinant selection rules could be refined to yield more compact reference spaces and faster convergence, for example by avoiding redundant determinants differing only by single or double excitations. Second, although UCCSD was employed here for its chemical interpretability, alternative entanglers, such as hardware-efficient circuits or qubit-adapted coupled-cluster circuits, could be explored to better match device constraints. Third, moving beyond small molecules in minimal basis sets to more complex systems, larger basis sets, and periodic materials will provide critical benchmarks for assessing the scalability and generality of CVQE.

Thanks to its modular structure, CVQE can also be seamlessly integrated with complementary quantum–classical strategies, including quantum subspace expansion, quantum self-consistent equation-of-motion methods~\cite{asthanaQuantumSelfconsistentEquationofmotion2023,kumar2023quantum,kwao2025generalized}, perturbative corrections, and even quantum machine learning or combinatorial optimization frameworks. In addition, CVQE remains compatible with other VQE paradigms, such as ADAPT-VQE~\cite{grimsleyAdaptiveVariationalAlgorithm2019} and ctrl-VQE~\cite{asthana2023leakage}. These features position CVQE as a versatile framework for accurate ground-state simulation in the NISQ era.

\section{Methods}

\subsection{Simulation Environment and Setup}

All simulations were performed using the \texttt{PennyLane} framework\cite{bergholmPennyLaneAutomaticDifferentiation2022} with the \texttt{lightning.qubit} backend for exact statevector evolution. For comparison, full configuration interaction (FCI) and selected configuration interaction (SHCI) energies were computed with \texttt{PySCF}\cite{sunPySCFPythonbasedSimulations} and \texttt{Arrow}\cite{holmesHeatBathConfigurationInteraction2016,sharmaSemistochasticHeatBathConfiguration2017,liFastSemistochasticHeatbath2018}, respectively.

Within the Born-Oppenheimer approximation, the molecular electronic Hamiltonian in second quantization takes the form,
\begin{equation}
H = \sum_{pq} h_{pq}\sum_\sigma a_{p,\sigma}^\dagger a_{q,\sigma} + \frac{1}{2} \sum_{pqrs} v_{pqrs} \sum_{\sigma,\sigma'} a_{p,\sigma}^\dagger a_{q,\sigma'}^\dagger a_{s,\sigma'} a_{r,\sigma}.
\end{equation} where $p,q,r,s$ denote spatial orbitals and $\sigma,\sigma'$ denote spin indices. One-body integrals $h_{pq}$ and two-body integrals $v_{pqrs}$ were obtained using \texttt{PennyLane}'s built-in functions with the STO-3G basis set. 

The fermionic Hamiltonian was mapped to qubits via the Jordan–Wigner transformation, with each spin orbital represented by a qubit (1 for occupied, 0 for unoccupied)\cite{mcardleQuantumComputationalChemistry2020}.
For example, the Hartree–Fock state for BeH$_2$ can be represented in \texttt{PennyLane} as an array
\begin{equation}
    \text{Hartree-Fock state: [1 1 1 1 1 1 0 0 0 0 0 0 0 0]},
\end{equation}
where qubits are ordered by increasing orbital energy, with $\alpha$-spin orbitals preceding $\beta$-spin orbitals.

\subsection{Molecular Systems}

We benchmark CVQE on BeH$_2$, H$_6$, and N$_2$ across a range of bond distances. 

\textbf{BeH$_2$} is a prototypical test molecule often used to probe the limitations of single-reference methods. Despite its apparent simplicity, stretching the Be–H bonds introduces static correlation effects that standard UCCSD approaches cannot capture. In our setup, BeH$_2$ has 7 spatial orbitals and 6 electrons, which map to 14 qubits. The molecule is modeled as a linear triatomic system in the H–Be–H configuration, with the hydrogen atoms placed symmetrically along the \(z\)-axis relative to the central beryllium atom. The bond distance \(d\) denotes the separation between each H and the central Be atom. The molecular geometry is given by
\begin{equation}
\text{BeH}_2: \quad \text{H} \ (0, 0, -d), \quad \text{Be} \ (0, 0, 0), \quad \text{H} \ (0, 0, d).
\end{equation}

\textbf{H$_6$} in a linear chain geometry is a canonical strongly correlated system. It consists of six hydrogen atoms equally spaced with bond length \(d\). The system contains 6 spatial orbitals and 6 electrons, requiring 12 qubits for mapping. As the H–H distances increases, the system undergoes a transition from a delocalized regime to a localized regime, leading to near-degeneracies among multiple Slater determinants. This property makes H$_6$ a stringent benchmark for testing quantum variational algorithms. Its geometry is defined as
\begin{equation}
\text{H}_6: \quad \text{H}_i \ (0, 0, i \cdot d), \quad i = 0, 1, ..., 5.
\end{equation}

\textbf{N$_2$} provides one of the most stringent tests for variational ansatz due to its triple-bond character, which demands simultaneous treatment of static and dynamic correlation during bond stretching. To balance correlation effects with computational cost, we treated N$_2$ in a reduced active space of 6 electrons in 6 spatial orbitals, mapped to 12 qubits. The nitrogen atoms are aligned along the \(z\)-axis with separation $d$:
\begin{equation}
\text{N}_2: \quad \text{N} \ (0, 0, 0), \quad \text{N} \ (0, 0, d).
\end{equation}

For all three molecules, the bond distance \(d\) was systematically varied from near-equilibrium to stretched regimes, thereby covering both weakly and strongly correlated domains.

\subsection{Trial State Preparation and Parameter Optimization}

At CVQE cycle $k$, the initial reference state is constructed as a superposition of Slater determinants from the set $\mathcal{S}^{(k)}$, with coefficients $\mathbf{c}$. This state was prepared using \texttt{PennyLane} built-in function \texttt{qml.Superposition}, which requires one auxiliary qubit. The circuit cost of this preparation scales linearly with the number of terms in the superposition. In the first CVQE cycle, the reference state is chosen as the Hartree-Fock determinant.

The reference state is then acted upon by a fixed entangling unitary, implemented using the unitary coupled-cluster ansatz with single and double excitations (UCCSD), $U_{\text{UCCSD}}$. Within the first-order Trotter approximation, the UCCSD unitary takes the form
\begin{equation}
\begin{aligned}
\hat U_{\text{UCCSD}}(\boldsymbol{\theta}) &=
\prod_{p>r} \exp\!\biggl\{ \theta_{pr}\bigl(c_p^\dagger c_r - \text{H.c.}\bigr) \biggr\} \\
&\quad \times
\prod_{p>q>r>s} \exp\!\biggl\{ \theta_{pqrs}\bigl(c_p^\dagger c_q^\dagger c_r c_s - \text{H.c.}\bigr) \biggr\}.
\end{aligned}
\end{equation}
Here $c^\dagger$ and $c$ denote fermionic creation and annihilation operators. The indices $r,s$ run over occupied orbitals and $p,q$ over virtual orbitals, defined relative to the Hartree--Fock reference state.
Fermionic operators were mapped to qubit operators using the Jordan-Wigner transformation\cite{seeleyBravyiKitaevTransformationQuantum2012}.
The parameters $\theta_{pr}$ and $\theta_{pqrs}$ are collectively denoted by $\boldsymbol{\theta}$.

In each cycle, both the ansatz parameters $\boldsymbol{\theta}$ and the determinant coefficients $\mathbf{c}$ are optimized using the vanilla gradient descent optimizer and the Cyclic Adamax (CAD) optimizer (see below), respectively.

We also observe that the optimal step size for gradient descent optimizer depends on both the molecular system and the bond length $d$. In practice, we employ step size of $0.1d$ or $0.05d$ for BeH$_2$, $0.5d$ for H$_6$ and N$_2$.

\subsection{Cyclic Adamax}

\textbf{Standard Adamax.} 
Adamax\cite{kingmaAdamMethodStochastic2017a} is a gradient-based optimizer designed to minimize high-dimensional loss functions by iteratively updating parameters according to gradient and momentum information. In the context of variational quantum algorithms, this loss function is typically the expectation value of the Hamiltonian, denoted in CVQE as $E(\mathbf{c})$.

Adamax maintains two running statistics of the gradient for each parameter $c_i$ at iteration $k$: 
(i) the first-moment estimate $m_i^{(k)}$ that captures momentum, and 
(ii) the max-norm estimate $u_i^{(k)}$ that tracks the largest observed gradient magnitude. 

The update rules are
\begin{align}
m_i^{(k)} &= \beta_1 m_i^{(k-1)} + (1-\beta_1) g_i^{(k)}, \\
u_i^{(k)} &= \max\!\big(\beta_2 u_i^{(k-1)}, \, |g_i^{(k)}|\big), \\
c_i^{(k+1)} &= c_i^{(k)} - \eta \, \frac{m_i^{(k)} / (1-\beta_1^k)}{u_i^{(k)} + \epsilon},
\end{align}
where $g_i^{(k)}$ is the gradient of the loss with respect to $c_i$ at iteration $k$, $\beta_1,\beta_2$ are decay rates, and $\eta$ is the learning rate. Here, $\epsilon$ is a small positive constant introduced to avoid division by zero.
Because the denominator uses the running maximum $u_i^{(k)}$, Adamax stabilizes step sizes even when gradients are noisy or unevenly scaled, which are common in near-term quantum simulations.

\medskip
\textbf{Cyclic restart mechanism.} 
Despite its stability, Adamax can stagnate when the optimization landscape becomes flat, as often occurs in barren plateaus. 
We address this challenge by introducing a \emph{cyclic restart mechanism}: after a fixed number of iterations (the restart period), the moment estimates $(m_i^{(k)}, u_i^{(k)})$ are reset or shrunk, discarding stale gradient information. This periodic restart encourages exploration of new descent directions and re-aligns the optimizer after reference expansion in CVQE.

\medskip
\textbf{Soft reset.} 
Instead of fully zeroing the moment vectors, a \emph{soft reset} scales them by a factor $\alpha \in [0,1]$:
\[
m_i^{(k)} \leftarrow \alpha \, m_i^{(k)}, 
\quad 
u_i^{(k)} \leftarrow \alpha \, u_i^{(k)}.
\]
When $\alpha=0$, this corresponds to a full reset; when $\alpha>0$, partial memory of past gradients is retained, enabling smoother adaptation. This flexibility allows the optimizer to adjust to different problem landscapes.

\medskip
\textbf{Adaptive reset.} 
In addition to fixed-period restarts, an \emph{adaptive reset} can be triggered dynamically when the gradient norm shows little variation over a sliding window, indicating stagnation on a plateau. In such cases, the optimizer automatically resets its moment estimates to regain progress.

\medskip
\textbf{Implementation in this work.} 
For all results reported here, we employed soft resets with $\alpha=0$ (full reset) at fixed intervals of 200 iterations, without adaptive resets. This setup proved effective across all systems studied, ensuring reliable convergence after reference-space expansion and enabling CVQE to consistently reach chemical accuracy. Nonetheless, future exploration of alternative restart strategies, including partial resets with nonzero $\alpha$ or adaptive resets, may further improve the convergence behavior of CVQE.

\subsection{Determinants selection}
In each CVQE cycle, after optimization, the trial state is sampled $n_{\mathrm{shots}}$ times in the computational basis. 
Any new Slater determinants observed with probability $p_i$ exceeding a threshold $p_{\mathrm{th}}$ are added to the reference set $\mathcal{S}$. 

Rather than using a fixed value, $p_{\mathrm{th}}$ is \emph{dynamically adjusted} during optimization to regulate the pace of determinants selection. Specifically, we set $p_{\mathrm{th}}$ proportional to the magnitude of the gradients of the variational parameters $\boldsymbol{\theta}$,
\begin{equation}
p_{\mathrm{th}} \propto \|\nabla_{\boldsymbol{\theta}} E\|.
\end{equation}
This strategy provides an adaptive schedule for determinants selection:  (i) in early stages, when gradients are large, only determinants with relatively high probability are admitted, prioritizing configurations associated with large coefficients;  (ii) as the energy converges and gradients decrease, the threshold automatically lowers, enabling the inclusion of determinants with smaller amplitudes.  

This gradual admission of new determinants enables CVQE to first capture dominant contributions efficiently and then refine the wavefunction with finer corrections, leading to more precise energy improvements.  

Newly admitted determinants are initialized with random small coefficients, scaled relative to the global norm of the coefficient gradients at that cycle, 
\begin{equation}
c_i^{\text{init}} \propto \omega\|\nabla_{\mathbf{c}} E\|,
\end{equation}
where $\omega$ is uniformly drawn from $(-1,1)$.

This adaptive initialization (i) prevents destabilizing the optimization by assigning coefficients that are commensurate with the current optimization scale, larger during the early stages and smaller as convergence proceeds; and (ii) promotes exploration of coefficient signs through randomization. 

\subsection{Determinants number control}
To balance accuracy with computational cost, CVQE impose an upper bound $n_{\mathrm{dets}}$ on the number of determinants retained in the reference set. 
When the selection procedure yields more candidates than this limit, pruning is applied: determinants with the smallest coefficients $|c_i|$ are discarded first, since their contribution to the wavefunction is negligible. They are then replaced by newly discovered determinants with larger expected importance, ensuring that the reference set remains both compact and effective.

The parameter $n_{\mathrm{dets}}$ thus serves as a practical control knob for CVQE. Increasing it systematically improves accuracy by enlarging the variational space, while decreasing it reduces quantum resource demands such as circuit depth and measurement overhead. This built-in tunability makes CVQE well suited to the resource constraints of near-term quantum devices, where both shot budgets and coherence times are limited.

\section*{Code and Data Availability}

The CVQE code and data supporting the findings of this study are openly available at the GitHub repository \href{https://github.com/hao-zhang-quantum/CyclicVQE}{https://github.com/hao-zhang-quantum/CyclicVQE}. The Cyclic Adamax (CAD) optimizer is provided as a separate file and can be readily reused for other applications.

\section*{Acknowledgments}

We thank Matthew Otten for helpful discussion. AA acknowledges NSF, award number 2427046, for support. 

\bibliographystyle{custom.bst}
\bibliography{bibliography}

\end{document}